\pdfoutput=1
\documentclass[11pt,usletter]{article}

\usepackage{amsthm}
\usepackage{jheppub}
\usepackage{slashed}

\usepackage{graphicx}
\usepackage{hyperref}
\usepackage{color}
\usepackage{epsfig}
\usepackage{amsfonts}
\usepackage{amssymb}
\usepackage{amsmath}
\usepackage{amsthm}
\usepackage{subfigure}
\usepackage{bm}
\usepackage{mathrsfs}
\usepackage{braket}
\usepackage{bbm}
\usepackage{cancel}
\usepackage{accents}

\def\ba{\begin{align}}
\def\ea{\end{align}}
\def\be{\begin{equation}}
\def\ee{\end{equation}}
\def\bea{\begin{eqnarray}}
\def\eea{\end{eqnarray}}
\usepackage[utf8]{inputenc}
\usepackage[english]{babel}

\def\Tr{{\rm Tr}}
\def\op{{\cal O}}

\def\vev#1{\langle{#1}\rangle}

\def\where{\quad {\rm where} \quad}

\def\and{\quad {\rm and} \quad}

\def\ra{\rightarrow}
\def\ie{{\rm i.e.\ }}

\def\CE{{\cal E}}

\def\CN{{\cal N}}

\begin{document}

\title{On thermalization in the SYK and supersymmetric SYK models}

\abstract{
The eigenstate thermalization hypothesis is a compelling conjecture which strives to explain the apparent thermal behavior of generic observables in closed quantum systems. Although we are far from a complete analytic understanding, quantum chaos is often seen as a strong indication that the ansatz holds true. In this paper, we address the thermalization of energy eigenstates in the Sachdev-Ye-Kitaev model, a maximally chaotic model of strongly-interacting Majorana fermions. We numerically investigate eigenstate thermalization for specific few-body operators in the original SYK model as well as its $\mathcal{N}=1$ supersymmetric extension and find evidence that these models satisfy ETH. We discuss the implications of ETH for a gravitational dual and the quantum information-theoretic properties of SYK it suggests.
}

\author[a]{Nicholas Hunter-Jones,}
\author[b]{Junyu Liu,}
\author[c,d]{and Yehao Zhou}

\affiliation[a]{Institute for Quantum Information and Matter,\\ California Institute of Technology, Pasadena, California 91125, USA}
\affiliation[b]{Walter Burke Institute for Theoretical Physics,\\ California Institute of Technology, Pasadena, California 91125, USA}
\affiliation[c]{Perimeter Institute for Theoretical Physics,\\ Waterloo, ON N2L 2Y5, Canada}
\affiliation[d]{Department of Physics \& Astronomy, University of Waterloo,\\ Waterloo, ON N2L 3G1, Canada}
\emailAdd{nickrhj@caltech.edu}
\emailAdd{jliu2@caltech.edu}
\emailAdd{yzhou3@perimeterinstitute.ca}

\maketitle
\flushbottom

\section{Introduction}
A deep understanding of the nature of thermalization and an approach to equilibrium in closed quantum systems remains to be seen. For a large class of interacting quantum many-body systems, the eigenstate thermalization hypothesis (ETH) \cite{DeutschETH,SrednickiETH} is an important conjecture which stands to help explain such a mechanism. The claim is that, in some large but isolated quantum mechanical system, the energy eigenstates look thermal, meaning that operators corresponding to physical observables have diagonal matrix elements given by microcanonical ensemble and off-diagonal elements suppressed by the entropy of the system. This conjecture gives us insight into the evolution of out-of-equilibrium states and is further supported numerically for few-body operators in specific strongly-correlated lattice models (see \cite{RigolETH} as well as \cite{ChaosETH} and references therein). The coincidence and difference between eigenstate expectation values and those in microcanonical ensemble provide a quantitative interpretation of how pure states become thermalized in a quantum chaotic system.

The mechanism of thermalization in a quantum theory is closely related to the statistical mechanics of black holes via the AdS/CFT correspondence, where spacetime geometries are dual to some specific quantum states in the boundary field theories. Understanding thermalization holographically \cite{Horowitz:1999jd} can help illustrate the types of bulk geometries dual to thermal states in the CFT and, more ambitiously, could shed light on black hole formation and evaporation. Moreover, the prototypical examples of eigenstate thermalization are quantum chaotic systems \cite{SrednickiETH,Srednicki98}. As black holes are both fast scramblers \cite{HaydenPreskill,SekinoSusskind} and maximally chaotic \cite{SSbutterfly}, explaining the emergence of thermalization from chaotic evolution will help us understand how black holes process quantum information \cite{ChaosChannels,ChaosDesign,ChaosRMT}.

The Sachdev-Ye-Kitaev (SYK) model \cite{Kitaev15,SachdevYe} is a concrete arena to address some of these questions. SYK is a quantum mechanical model of $N$ strongly-interacting Majorana fermions with all-to-all random couplings. The model is solvable at large $N$ and possesses many intriguing features indicative of a gravitational dual, such as an emergent conformal symmetry at low energies and a zero temperature entropy \cite{Kitaev15,MS_SYK}. The theory is also maximally chaotic in the sense that an explicit calculation of the out-of-time ordered correlation function \cite{Kitaev15,MS_SYK,Polchinski:2016xgd} shows a Lyapunov growth which saturates the chaos bound \cite{MSSbound}, a seemingly unique feature of gravitational theories \cite{SSstringy} as well as CFTs with a gravitational dual \cite{ChaosCFT}. The low energy description of the theory is given by a Schwarzian action which also captures dilaton gravity in AdS$_2$ \cite{Jensen:2016pah,SYKbulk}. Finally, SYK exhibits both random matrix universality in the spectrum \cite{You16,Garcia16} as well as random matrix behavior in the late-time dynamics of the model \cite{BHRMT16}. Many of these features are present in generalizations of SYK that have been discussed, including a supersymmetric extension of the model.

In this paper, we will discuss ETH in SYK and SYK-like models. We will check the original SYK model and the $\mathcal{N}=1$ supersymmetric generalization of SYK given by Fu, Gaiotto, Maldacena, and Sachdev \cite{SUSY_SYK}. By exact diagonalization, we numerically compute the matrix element in the energy eigenstates and verify that these models satisfy ETH. Our numerics quantitatively illustrate that pure states in these holographic models can thermalize.

There has already been some discussion in the literature of thermalization in SYK, both looking at ETH as well as considering more dynamical questions. It has been shown \cite{Magan15} that ETH is satisfied in the free fermion analog of SYK. \cite{Eberlein:2017wah} discussed a quench setup in the SYK model and analytically showed that the system rapidly thermalizes in a certain limit. Diffusion and spread of entanglement have also been discussed in a generalization of SYK with spatial locality \cite{Gu:2016oyy,Gu:2017njx}. Recently, \cite{Kourkoulou:2017zaj} showed that specific pure states in the SYK model are thermal in the large $N$ limit. Most relevant to this work, \cite{Sonner17} investigated ETH in the complex fermion version of SYK model and found evidence for eigenstate thermalization. While the model with complex fermion version of SYK \cite{SachdevBH,FuSachdev,Davison:2016ngz} shares many of the same features, albeit with some subtle differences \cite{ComplexSYK}, one is still inclined to believe that thermalization and ETH will hold much in the same way for both Majorana SYK and its supersymmetric extension. We investigate ETH explicitly in these models and, for some few-body operators, check the ansatz for both diagonal and off-diagonal terms, and investigate the Gaussianity of fluctuations.


This paper is organized as follows: In Section \ref{sec:setup} we review the SYK model and its supersymmetric generalization, then discuss the basics of the eigenstate thermalization hypothesis. In Section \ref{sec:num} we present a numerical investigation, by exact diagonalization, of the matrix elements of specific operators evaluated in pure states. We read off the corresponding behavior of the diagonal and off-diagonal elements. We conclude in Section \ref{sec:con} and discuss the implications for thermal properties of the bulk and quantum information aspects of SYK. In Appendices \ref{app:PHsym} and \ref{app:UI} we mention two technical points, including a constraint from particle-hole symmetry in the SYK model, and the issue of unitary averaging of distributions when testing Gaussianities of fluctuations.

\section{Overview}
\label{sec:setup}
\subsection{SYK and supersymmetric SYK}
Before we delve into a numerical investigation, we first review the quantum mechanical models we wish to consider. The Hamiltonian of the ($q$-point) SYK model \cite{Kitaev15,MS_SYK} is given by
\begin{equation}
H = (i)^{q/2} \sum_{i_1<i_2<\ldots<i_q} J_{i_1 i_2\ldots i_q} \psi^{i_1}\psi^{i_2}\ldots \psi^{i_q}\,,
\end{equation}
where $\psi^i$ denote $N$ Majorana fermions, satisfying the anticommutation relation $\{\psi^i, \psi^j \}=\delta^{ij}$. The system has all-to-all $q$-body interactions with random couplings $J_{i_1 i_2\ldots i_q}$ with mean and variance
\begin{equation}
\big\langle J_{i_1 i_2\ldots i_q} \big\rangle=0\,, \qquad \big\langle J_{i_1 i_2\ldots i_q}^{2} \big\rangle =\frac{J_{\rm SYK}^2 (q-1)!}{N^{q-1}}\,,
\end{equation}
i.e.\ we disorder average over independent Gaussian random variables $J_{i_1 i_2\ldots i_q}$, and $J_{\rm SYK}$ is a positive constant. The model is exactly solvable in the limit $1\ll \beta J_{\rm SYK} \ll N $, where an emergent conformal symmetry can be used to compute the correlation functions of the theory. A calculation of the out-of-time-ordered four point function of the theory \cite{MS_SYK,Kitaev15} shows that the Lyaponov exponent, governing the time-dependent growth of $1/N$ corrections, saturates the chaos bound $\lambda=2\pi/\beta$ at large $N$. The conformal symmetry is spontaneously and explicitly broken and one can obtain a low-energy description of the model given by the Schwarzian, capturing the dynamics of the pseudo-Goldstone mode. The Schwarzian theory can also be understood as the low-energy dilaton gravity description of the two-dimensional bulk dual of \cite{SYKbulk,KitaevSuh}.

Interesting generalizations to supersymmetric versions of the SYK model were originally given in \cite{SUSY_SYK}, which have been further studied in \cite{Murugan:2017eto,Yoon:2017gut,Peng:2017spg,Li:2017hdt,Kanazawa:2017dpd,Peng:2016mxj}. Here we just discuss the $\mathcal{N}=1$ generalization. In the $(2q-2)$-point $\mathcal{N}=1$ supersymmetric model, one can construct an $\mathcal{N}=1$ supercharge $Q$ as a $q$-body Majorana interaction. The Hamiltonian is then given by the square of the supercharge
\begin{equation}
H=Q^2\,, \where Q=i^{(q-1)/2} \sum_{i_1<i_2<\ldots<i_q} C_{i_1 i_2\ldots i_q} \psi^{i_1}\psi^{i_2}\ldots \psi^{i_q}\,,
\end{equation}
and again we have random couplings $C_{i_1 i_2\ldots i_q}$ with mean and variance
\begin{equation}
\big\langle C_{i_1 i_2\ldots i_q} \big\rangle=0\,, \qquad \big\langle C_{i_1 i_2\ldots i_q}^{2} \big\rangle =\frac{J_{\CN=1}^{2}(q-1)!}{{{N}^{q-1}}}\,.
\end{equation}
Just as above, $C_{i_1 i_2\ldots i_q}$ is an antisymmetric Gaussian random tensor and $J_{\CN=1}$ is a constant. The bosonic operators $b^i$ of the theory, supersymmetric partners of $\psi^i$, are given by the application of the supercharge as $b^i = Q\psi^i$. This model shares many of the intriguing large $N$ features observed in the Majorana model, such as chaotic correlation functions, zero-temperature entropy, and an emergent superconformal symmetry that is spontaneously and explicitly broken, giving rise to a Schwarzian-like low-energy action \cite{SUSY_SYK}.

In this paper we will only consider the simplest non-trivial interactions in the SYK models, meaning we set $q=4$ in the SYK model and consider $q=3$ supercharges in the $\CN=1$ supersymmetric model. We will also choose $J_{\rm SYK} = J_{\CN=1} = 1$ without loss of generality, as the coupling can simply be regarded as a rescaling of energy.

As our best understood examples of AdS/CFT involve string theory in AdS, with a supergravity description at low-energies, the supersymmetric extension of SYK might be helpful in constructing a more concrete gravity dual in two dimension AdS or understanding how the dilaton gravity might be UV completed. With this in mind, we should note the similarities and differences for thermalization in the original model compared to its supersymmetric extension.

\subsection{Eigenstate thermalization}
ETH is a statement about the thermal nature of the energy eigenstates in a closed quantum system \cite{DeutschETH,SrednickiETH}, where the thermal behavior emerges from pure states for some generic local operators. The claim is that the matrix elements of an operator $\mathcal{O}$, evaluated in the basis of energy eigenstates, should satisfy the following:
\begin{equation}
\left\langle  m|\op| n \right\rangle =\bar{\op}(\bar{E})\delta_{mn} + e^{-S(\bar{E})/2} f_\op (\bar{E},\omega ) R_{mn}\,,
\label{eq:ETH}
\end{equation}
where $\bar{E}=(E_m+E_n)/2$, and $\omega=E_m-E_n$. The function $\bar \op(\bar E)$ is the average of the observable $\op$ in the microcanonical ensemble, $\bar \op = \Tr (\rho_{\rm mc} \op)$, and a smooth function of $\bar E$. Moreover, the function $f_\mathcal{O}(\bar{E},\omega)$ is a smooth function of $\bar{E}$ and $\omega$, $S(\bar{E})$ is the entropy of the system, and $R_{mn}$ is a random variable with zero mean and unit variance. ETH states that in a system with many degrees of freedom, the diagonal matrix elements dominate, while the off-diagonal terms are suppressed by the entropy of the system. The distribution of $R_{mn}$ could be real or complex, depending on the symmetry of the system we consider and the basis we choose. In either case complex conjugate constrains the matrix elements of $R_{mn}$ and the function $f_\mathcal{O}$.

ETH, as formulated above, can help explain the emergence of thermal expectation values of certain operators in the time evolution of quantum many-body systems. Following the discussion in \cite{Srednicki98,ChaosETH}, we prepare a pure state $\ket{\psi}$ in an isolated quantum system with Hamiltonian $H$ and energy eigenstates $\ket{n} $, where the time evolved state is given as
\begin{align}
\ket{\psi (t)}=\sum\limits_{n}{{{c}_{n}}{{e}^{-i{{E}_{n}}t}}\left| n \right\rangle }\,,
\end{align}
with coefficients $c_n$ given by $c_n =\braket{n|\psi_i} $. The expectation value of any operator $\op$ in the time-evolved state is
\begin{align}
\braket{\op (t)} = \sum_n |c_n|^2 \op_{nn} + \sum_{m\ne n} c_m^* c_n e^{i(E_m-E_n)t} \op_{mn}\,,
\end{align}
where we denote matrix elements of the operator $\op_{mn} = \braket{m|\op|n} $. At late times, the off-diagonal terms are weighted with a dephasing exponential factor, which causes them to decay. More precisely, we can take an infinite time average and find
\begin{align}
\braket{\op (t)} \approx \sum_n |c_n|^2 \op_{nn}\,,
\end{align}
assuming that there are no, or at least a nonextensive number of, degeneracies (\ie not scaling with the system size). This late-time value of the operator is often called its expectation value in the diagonal ensemble. One should note that the leading term in this expectation value at late time is constant and only depends on the construction of the initial state $c_n$. On the other hand, the ensemble average in the microcanonical ensemble is
\begin{equation}
\bar \op (E)=\frac{\sum_{n\in \Delta E} \op_{nn}}{N_{\Delta E}}\,,
\end{equation}
where we sum over the states in some narrow window of energies near $E$, \ie sum over $n$ such that $|E-E_n|<\Delta E$, and $N_{\Delta E}$ denotes the number of states in that window. Now assuming that the initial state is prepared in a narrow energy window around $E$, according to the assumption Eq.~\eqref{eq:ETH}, $\bar{\mathcal{O}}(E)$ is a smooth continuous function and thus near the energy window $E$ the matrix element $\mathcal{O}_{nn}\approx\bar{\mathcal{O}}(E)$. Thus we have
\begin{equation}
\left\langle \mathcal{O}(t) \right\rangle \approx \sum\limits_{n}{{{\left| {{c}_{n}} \right|}^{2}}\bar{\mathcal{O}}(E)}=\bar{\mathcal{O}}(E)\sum\limits_{n}{{{\left| {{c}_{n}} \right|}^{2}}}=\bar{\mathcal{O}}(E)\,,
\end{equation}
which shows the emergence of thermalization in a system that satisfies ETH. All we needed to assume was that the diagonal and microcanonical distributions were sufficiently narrow. More precisely, Srednicki gave a criteria \cite{Srednicki95} for the smallness of $\Delta E$ relative to the smooth function of diagonal elements, which we identify with microcanonical average of the operator. If the energy fluctuations are subextensive, then we conclude that the microcanonical and long-time averages agree, and thus the system thermalizes.

Clearly, ETH cannot hold for all possible operators. In fact, one can construct operators to trivially violate ETH in all possible quantum systems, \ie projection operators or operators constructed from the Hamiltonian. However, ETH is still far from trivial; these artificial operators are nonlocal and thought not to correspond to real physical observables. It is often claimed that ETH should hold for few-body operators that are sufficiently local. Interestingly \cite{Grover15}, it was argued that ETH should hold for operators with spatial support up to half of the system size. ETH was first observed numerically in a bosonic lattice model \cite{RigolETH}, and further been observed in a variety of strongly-interacting lattice models, including spin chains, fermionic systems, etc. \cite{SantosLoc,RigolGap,SantosOnset,RigolQuen,ETH_TFIM}, including checks in every eigenstate \cite{KimTest}.

Another subtlety arises when discussing ETH in a system with disorder. In a disordered system the eigenvalues and eigenstates are not fixed but instead depend on the specific realization of disordered couplings. But ETH still makes sense in the following way. For a given energy $E$ and some number of disorder realizations, the energy eigenvalues of those realizations will vary around $E$. For a given energy window $\Delta E$, we can look at the energy eigenstates for all possible eigenvalues that are in the range $[E-\Delta E,E+\Delta E]$ and compute the statistics of those eigenstates (for instance, for fixed $\bar{E}$ and $\omega$, or equivalently, $E_m$ and $E_n$, one considers the eigenvalues in the range $[E_m-\Delta E,E_m-\Delta E]$ and $[E_n-\Delta E,E_n-\Delta E]$ for a given number of random realizations). For ETH in disordered systems, the eigenstates $|m\rangle$ and $|n\rangle$ are not fixed but rather ensembles of eigenstates in an energy window $\Delta E$. Similar issues are discussed and addressed in \cite{ChaosETH,ETHgauss}.


There is a long history between ETH and systems which exhibit quantum chaos\footnote{We should emphasize that quantum chaos is still without a precise definition. Yet there are universal phenomena one expects in a quantum chaotic system, such as having the spectral statistics of a random matrix and a chaotic decay of out-of-time-ordered correlation functions. It is the first notion which is more traditionally associated with ETH. But a deeper relation between the two symptoms of chaos might exist. For instance, the SYK model has both a maximal chaos exponent at large $N$ \cite{Kitaev15,MS_SYK} and exhibits random matrix behavior in its spectral statistics and late-time dynamics \cite{You16,Garcia16,BHRMT16}.} (as nicely reviewed in \cite{ChaosETH}). If a chaotic system is highly random it sometimes suffices to consider a Hamiltonian given by a random matrix. In this case, without assuming the initial state is narrow, we can often directly take $\bar{\mathcal{O}}(E)$ outside of the summation. Given the maximal chaos and late-time random matrix behavior in SYK-like models, we expect that they should satisfy ETH. For a discussion of the random matrix behavior of SYK, see \cite{BHRMT16,ChaosRMT}; for supersymmetric models, see \cite{Li:2017hdt,Kanazawa:2017dpd} and \cite{ChaosSUSY}. Recent work also provides evidence which supports this conjecture; for instance, \cite{Kourkoulou:2017zaj} shows that aspecific low-energy pure state correlators are equal to thermal correlators at large $N$, while \cite{Sonner17} gave numerical evidence that the complex SYK model satisfies ETH. In the following section, we will show precisely how ETH is satisfied for the Majorana SYK model as well as its supersymmetric extension.

\section{Numerics for Thermalization}
\label{sec:num}
\subsection{Setup}
In this part we will describe our numerical results. We study the system by exact diagonalization. For SYK model and supersymmetric SYK model, we use the Clifford algebra representation of Majorana fermions, where we employ the standard Pauli matrices as
\begin{equation}
\gamma_{2k} =\frac{1}{\sqrt{2}} \left( \prod_{i =1}^{N_d-1} Z_i \right) Y_{N_d}\,, \qquad
\gamma_{2k-1} = \frac{1}{\sqrt{2}} \left( \prod_{i =1}^{N_d-1} Z_i \right) X_{N_d}\,,
\label{eq:gamma}
\end{equation}
with $N_d=N/2$ and $k\in\{1,2,\ldots,N_d\}$, satisfying the anticommutation relation $\{ \gamma_i, \gamma_j\}  = \delta_{ij}$.

To test ETH, we must restrict to a specific class of operators. One can consider the particle number operator,
\begin{equation}
n_k = c_k^\dagger c_k\,,
\end{equation}
where $k\in \{1,2,\ldots,N_d\}$, and $c_k$ is a complex fermion. We may write Majoranas as
\begin{equation}
\psi^{2k} = \frac{1}{\sqrt{2}}\big( c_k+c_k^\dagger \big)\,, \qquad \psi^{2k-1} = \frac{i}{\sqrt{2}} \big(c_k-c_k^\dagger\big)\,.
\end{equation}
The particle number operator $n_k$ could also be written as
\begin{equation}
n_k = \frac{1}{2}(S_k+1)\,,
\end{equation}
where $S_k$ is defined as \cite{Kourkoulou:2017zaj}
\begin{equation}
S_{k}=2i{{\psi }^{2k-1}}{{\psi }^{2k}}\,.
\end{equation}
This operator always have the eigenvalues $s_k=\pm 1$. For each $k$, half of the $2^{N_d}$ dimensional eigenspace has the eigenvalue $+1$, while the rest has the eigenvalue $-1$. Thus, one can write down the common eigenvector for all $k$, with a given series of $s_k$ \cite{Kourkoulou:2017zaj}
\begin{equation}
S_{k} \ket{B_s} = s_k \ket{B_s} \quad\text{for all } k\,,
\end{equation}
where, in a slight abuse of notation, the subscript denotes $s\equiv \vec s = \{s_1,\ldots,s_k\}$. The $2^{N_d}$ states $\ket{B_s}$ form a complete basis of the Hilbert space (and in fact, are unit vectors in the representation Eq.~\eqref{eq:gamma}). Thus, we may write
\begin{equation}
S_k = \sum_s \lambda_s  \ket{B_s}\!\bra{B_s}\,,
\end{equation}
where half of the $\lambda$'s in the sum are $+1$ and other half are $-1$.
As the operator $S_k$ has no thermal expectation value in the diagonal terms and we only have random fluctuations in both the diagonal and off-diagonal elements. Note that the diagonal terms for $n_k$ are non-trivial and fluctuate, except for when the particle number is fixed by symmetry as we discuss in Appendix \ref{app:PHsym}.

To further test ETH, another natural 2-body operator to consider is the hopping operator
\begin{align}
h_{k_1 k_2} = c_{k_1}^\dagger c_{k_2} + c_{k_2}^\dagger c_{k_1}\,.
\end{align}
Note that for $k_1\ne k_2$, this operator has vanishing expectation value in the microcanonical ensemble $\Tr ( \rho_{\rm mc} h_{k_1 k_2} ) = 0$. So agreement with ETH will mean the diagonal elements are close to zero.

In the following sections, we both check ETH for $n_k$ and $h_{k_1,k_2}$. For simplicity, we only consider models with even $N_d$ in performing our numerics. In both the SYK model and its supersymmetric extension, fermion number parity is conserved, and thus we can focus on one single parity sector to study the ETH. Without loss of generality, we will simply look at the even parity sector.

\begin{figure}
\centering
\includegraphics[width=0.48\textwidth]{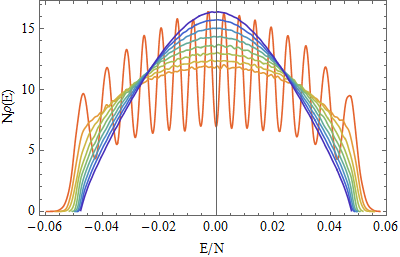} \quad
\includegraphics[width=0.48\textwidth]{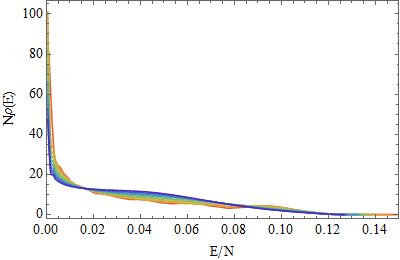}
\caption{\label{rho} We plot the density of states, on the left for the SYK model and on the right for the $\mathcal{N}=1$ supersymmetric SYK model, for $N =$ 12, 16, 20, 24 and 28 Majoranas, where we take 25600 $(N=12)$, 6400 $(N=16)$, 1600 $(N=20)$, 400 $(N=24)$, and 100 $(N=28)$ realizations of disorder. The above data is collected from the full spectrum.}
\end{figure}

We quickly note mention some aspects of the expectation values of particle number operators specific to the Majorana SYK and its supersymmetric generalization. These models have a particle-hole symmetry, where the antiuntiary operator $P$ commutes with the Hamiltonian $H$ \cite{You16,BHRMT16}. $P$ maps a state with fermion number $Q$ to $N_d-Q$, or equivalently sends the filling from $\nu$ to $1-\nu$ when acting on a pure state, where $\nu = n_f/N_d$. If $N \text{ mod 8} =0$ where $P^2=1$, then $P$ maps each energy eigenstate to itself, which means we have the constant filling $\nu=1/2$ for all possible energy eigenstates. In other cases with even $N_d$ we are considering, the particle-hole symmetry is non-trivial and enforces a degeneracy in each parity sector. For a more details on particle-hole symmetry and the constraint described above, see Appendix \ref{app:PHsym}. Randomly looking at energy eigenstates states from enlarged eigenspace, the particle number expectations should be fluctuating around $1/2$. Thus, when calculating diagonal elements we will avoid the $N \text{ mod 8} =0$ case.

\begin{figure}
\centering
\includegraphics[width=1\textwidth]{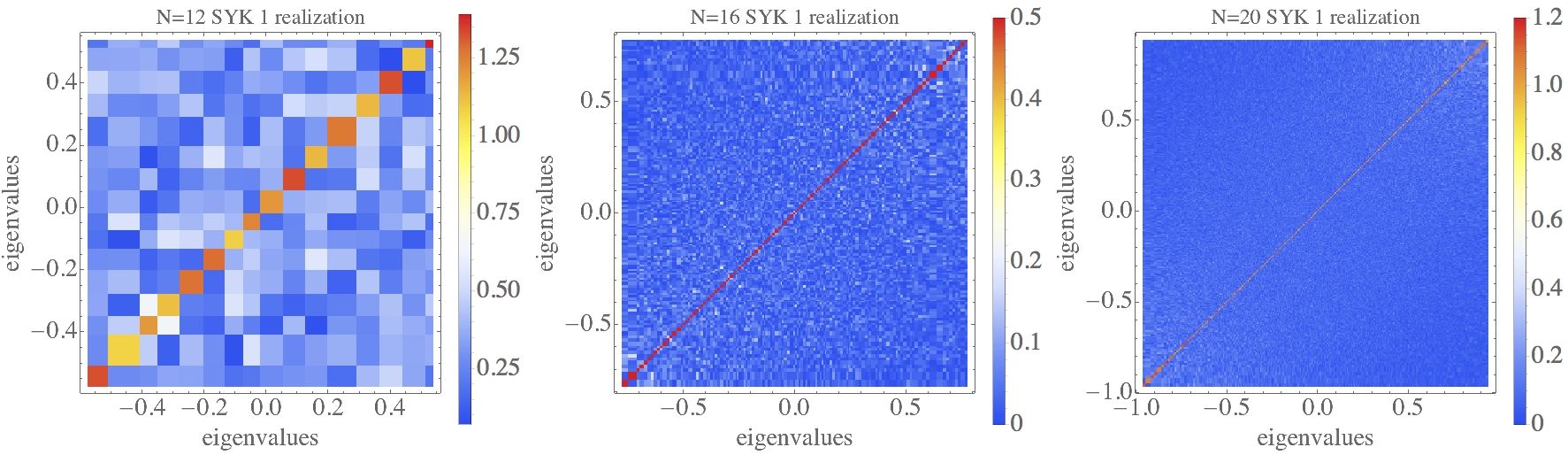}\\
\includegraphics[width=0.68\textwidth]{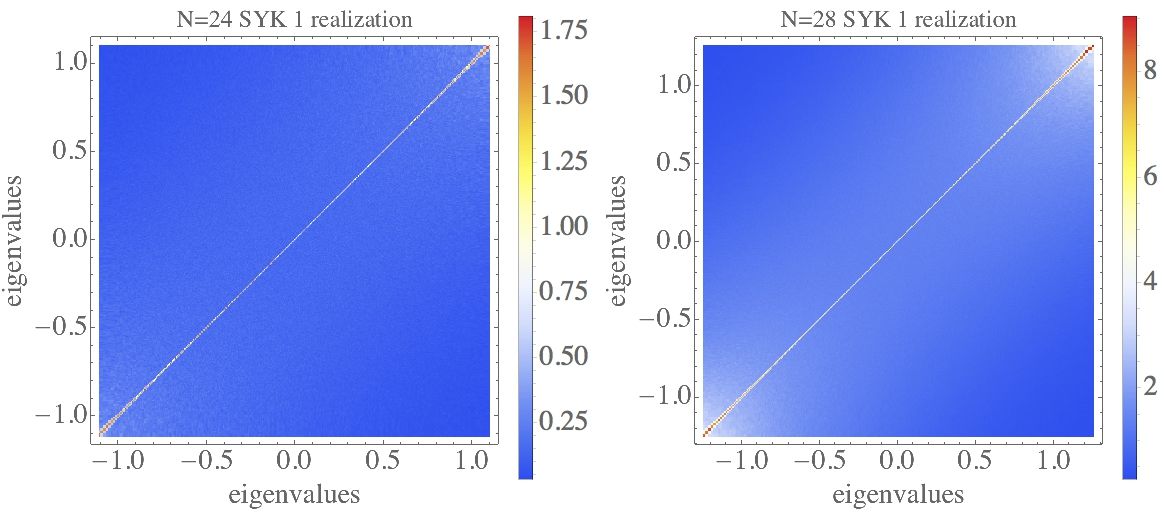}
\caption{Eigenstate thermalization density plots for the particle number operator in the original SYK model. We choose a single realization of the model with $N=$ 12, 16, 20, 24, and 28. The horizontal and vertical axes denote the energy eigenvalues for the corresponding matrix elements, while the value of the density is the complex norm of the matrix element. Given the rapid growth in the number of eigenvalues, we downsample the $N=24$ and $N=28$ data to $512\times512$. Considering the possible degeneracies, we average the doubled data points with the same energy eigenvalues.} \label{syk_cc}
\end{figure}

\subsection{Density of states}
Before considering the matrix elements of specific operators projecting on the spectrum and discussing the thermalization of the eigenstate, in Figure \ref{rho} we plot the density of states for SYK and $\mathcal{N}=1$ supersymmetric SYK model for completeness (those results and related numerics are previously obtained in, for instance, \cite{MS_SYK,Li:2017hdt}). The spectral density of SYK has been further studied in \cite{BHRMT16,Garcia16,Garcia17}. The density of states are obtained for multiple realization of randomness in the corresponding Hamiltonian. In the (relatively) large $N$, the distribution of eigenvalues for SYK model will be more closer to a combination between Gaussian distribution (by the central limit theorem) and the Wigner's semicircle, while in the supersymmetric model, the supercharge is Gaussian-like, thus the distribution of energy eigenvalues will approach the square of Gaussian randomness: the Marchenko-Pastur distribution in a large $N$ limit of the one point distribution in the Wishart-Laguerre ensemble (see \cite{Li:2017hdt} for more details).

\begin{figure}
\centering
\includegraphics[width=1.0\textwidth]{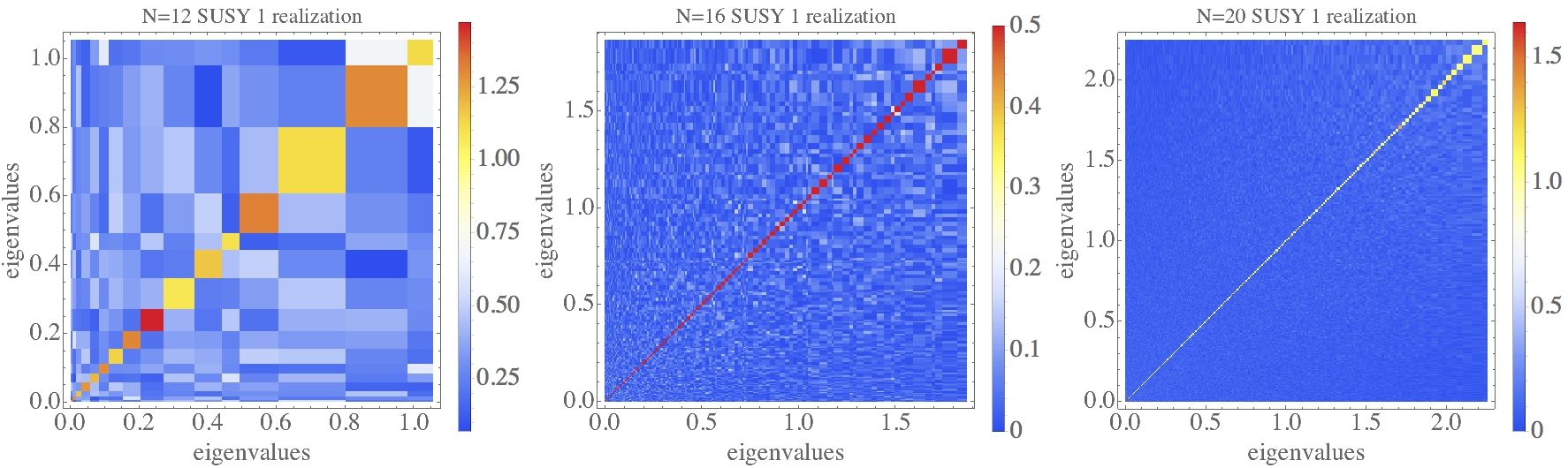}\\
\includegraphics[width=0.68\textwidth]{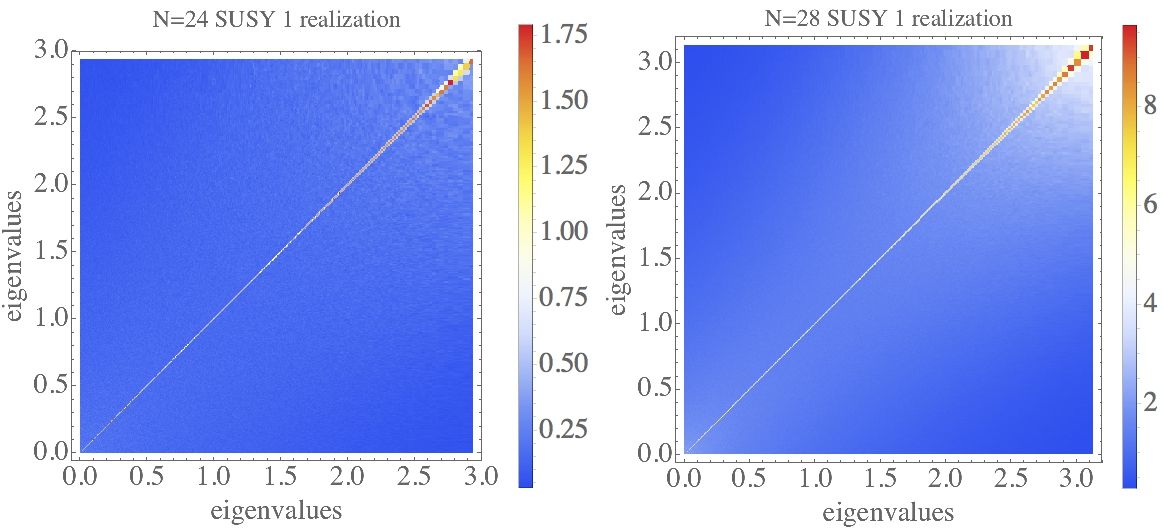}
\caption{Eigenstate thermalization density plots for the particle number operator in the supersymmetric SYK model. The setup is the same as described in Figure \ref{syk_cc}.} \label{susy_cc}
\end{figure}

\subsection{Direct checks of ETH}
\begin{figure}
\centering
\includegraphics[width=1.0\textwidth]{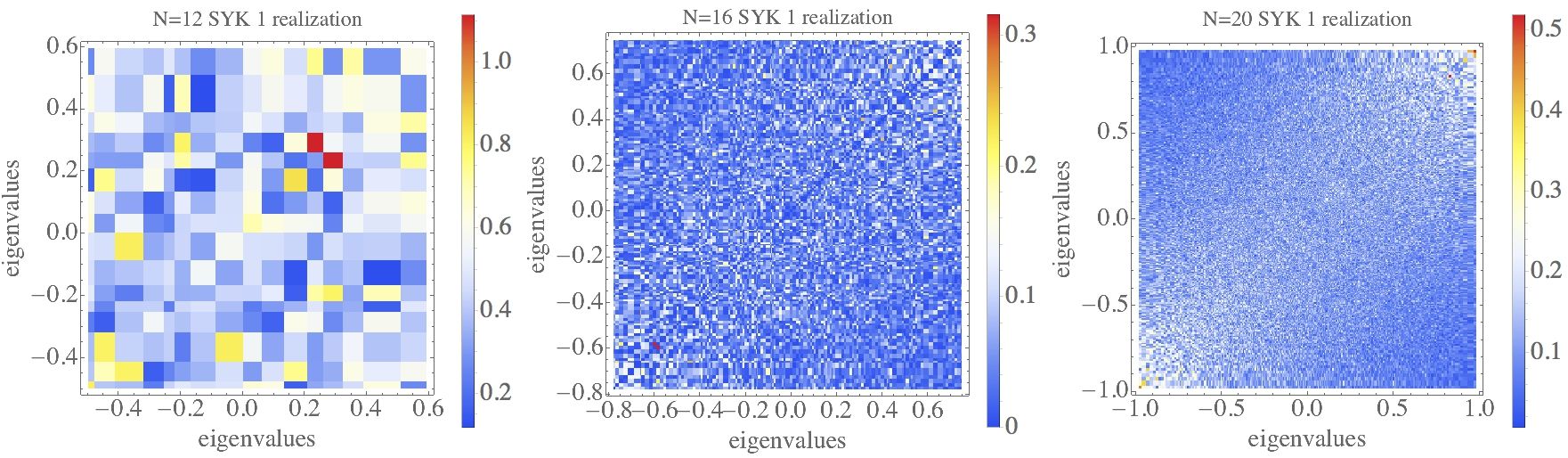}
\includegraphics[width=0.68\textwidth]{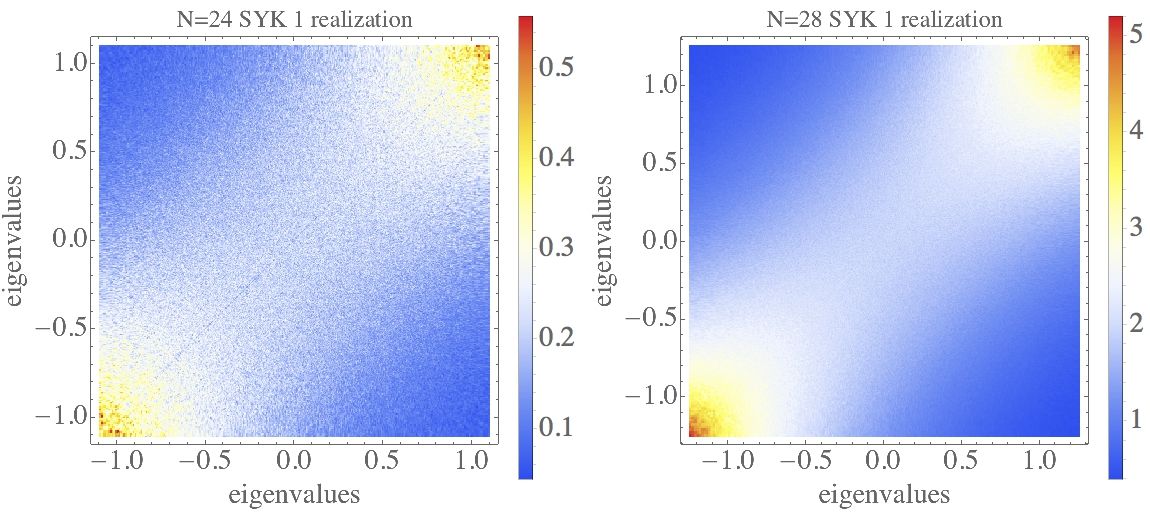}
\caption{Eigenstate thermalization density plots for single hopping operators in the original SYK model, with the same setup as described in Figure \ref{syk_cc}. Note that the microcanonical value of the hopping operator is zero, so agreement with ETH is simply the appearance of the off-diagonal Gaussian fluctuations.} \label{syk_hop}
\end{figure}

\begin{figure}
\centering
\includegraphics[width=1.0\textwidth]{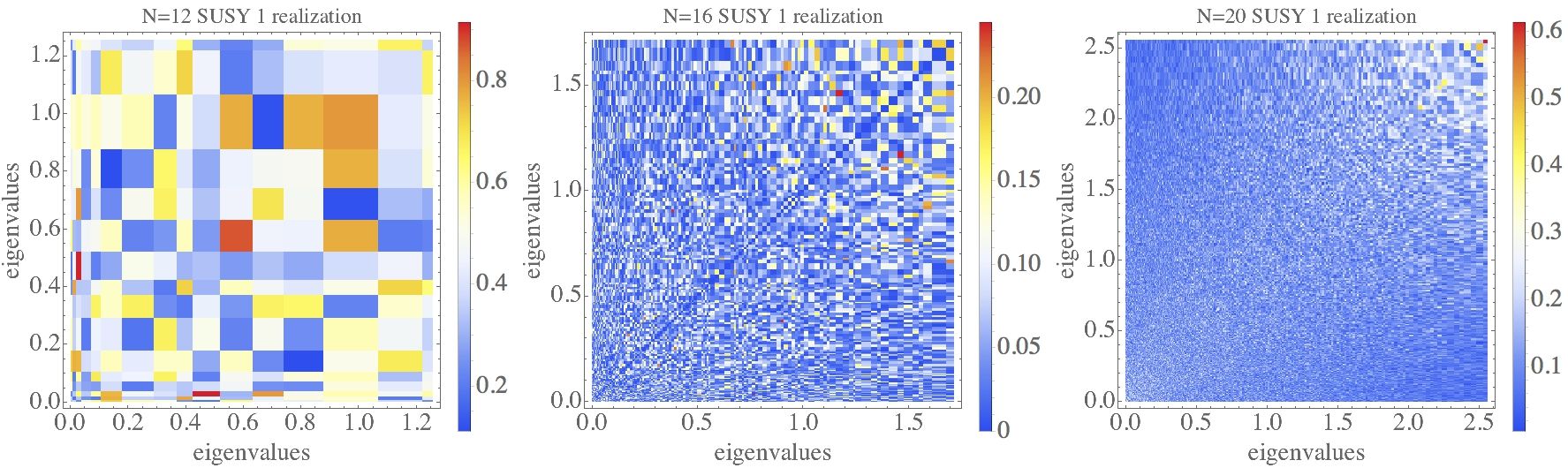}
\includegraphics[width=0.68\textwidth]{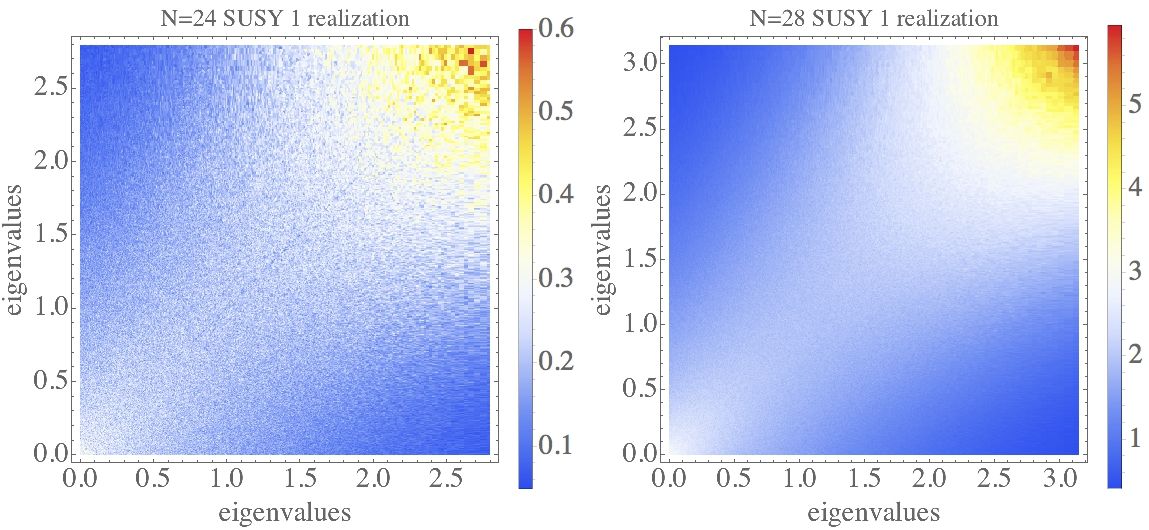}
\caption{Eigenstate thermalization density plots for single hopping operators in the supersymmetric SYK model. The setup is the same as described in Figure \ref{syk_cc}.} \label{susy_hop}
\end{figure}

Now we come to direct checks of ETH in SYK models. First, we make density plots of the matrix elements of number operators and hopping operators projecting onto the energy eigenstates in both models. The evidence of eigenstate thermalization is apparent even in a single random realization of disorder, where we do not need to take a thermal average. For the particle number operators, as shown in Figure \ref{syk_cc} and Figure \ref{susy_cc} for SYK and supersymmetric model respectively, the density plot clearly shows the diagonal dominance in the matrix elements $\op_{mn}$, while the off-diagonal terms are suppressed by the entropy of the system, with small random fluctuations. However, if we consider hopping operators, as displayed in Figure \ref{syk_hop} and Figure \ref{susy_hop}, there is no $\bar{\op}(\bar{E})$ contribution in the microcanonical ensemble thus confirmation with ETH is simply the appearance of small random fluctuations in the matrix elements, as observed.

\begin{figure}
\centering
\includegraphics[width=0.8\textwidth]{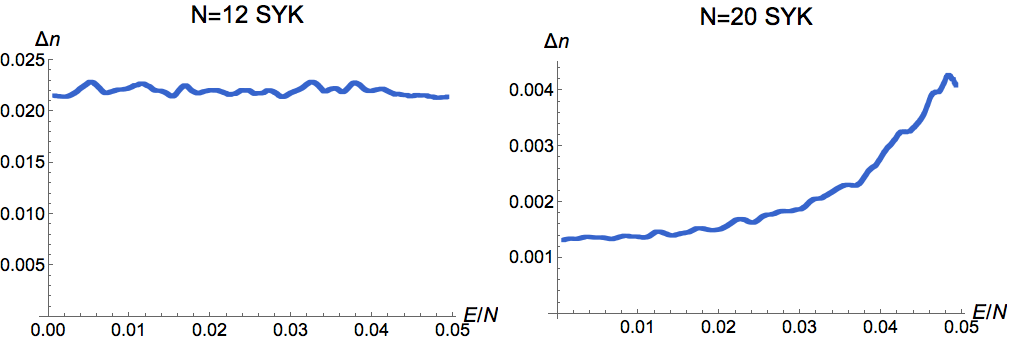}\\
\includegraphics[width=0.8\textwidth]{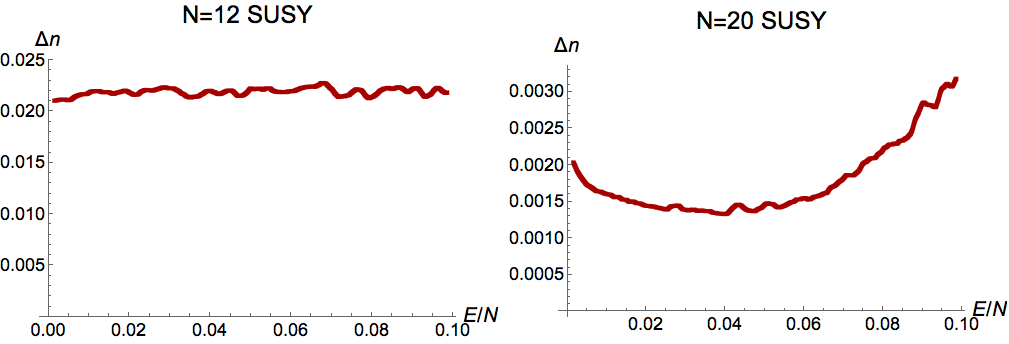}
\caption{The relative difference between the microcanonical ensemble predictions and the diagonal ensemble predictions in the SYK model (above) and the supersymmetric SYK model (below) as a function of energy $E/N$. We choose multiple realizations for $N=12$ (left, 20000 realizations) and $N=20$ (right, 1000 realizations) models respectively. We consider the energy window $\Delta E=0.01$ in both cases. In this plot, we totally collect 1001 data points and in order to smooth the random fluctuation we use the moving average for 30 nearest energy data points. The remaining random fluctuation may due to the finiteness of the number of random generalizations.} \label{ratio}
\end{figure}

Another direct investigation of ETH is to evaluate the difference between the eigenstate projection and the expectation value from the microcanonical ensemble \cite{RigolBreak}, \ie the difference between diagonal and microcanonical ensembles. To be concrete, we consider
\begin{equation}
\Delta \op = \frac{\sum_{n \in \Delta E} \big| \op_{nn} - \bar{\op}(E) \big| }{\sum_{n \in \Delta E} \op_{nn}}\,,
\end{equation}
where we sum over all energies (\ie eigenvalue labels $n$) in the energy window $[E-\Delta E,E+\Delta E]$, and $\bar{\mathcal{O}}(E)$ is given as the microcanonical ensemble average for the energy $E$. Specifically, we consider the particle number operator $\mathcal{O}=n_k$. As explained previously, the diagonal terms are strictly $1/2$ for $N \text{ mod } 8=0$ for each realization of disorder parameter, thus we specifically consider $N=12$ and $N=20$ in Figure \ref{ratio} for SYK and its supersymmetric extension.
The results indicate thermalization of eigenstates as the relative difference is very small and becomes further suppressed as we increase the number of fermions. The difference between the microcanonical and diagonal predictions depends on the energy (or the effective temperature in those models). But in the parameter range we are testing, we conclude that the predictions from the microcanonical and diagonal ensembles are very close, although there are still random fluctuations due to the finiteness of the sample size.

\subsection{Diagonal terms}
Some checks of eigenstate thermalization can also made in the diagonal terms $\mathcal{O}_{nn}$. For the single fermion particle number $n_k$, we should observe fluctuations around the microcanonical value $1/2$ (except for $N \text{ mod } 8=0$ where there are no fluctuations as the particle number is fixed by symmetry). In the other cases, the system may still have a degeneracy due to the internal symmetry in each parity sector and thus we only choose one eigenstate in each energy level.

\begin{figure}
\centering
\includegraphics[width=0.92\textwidth]{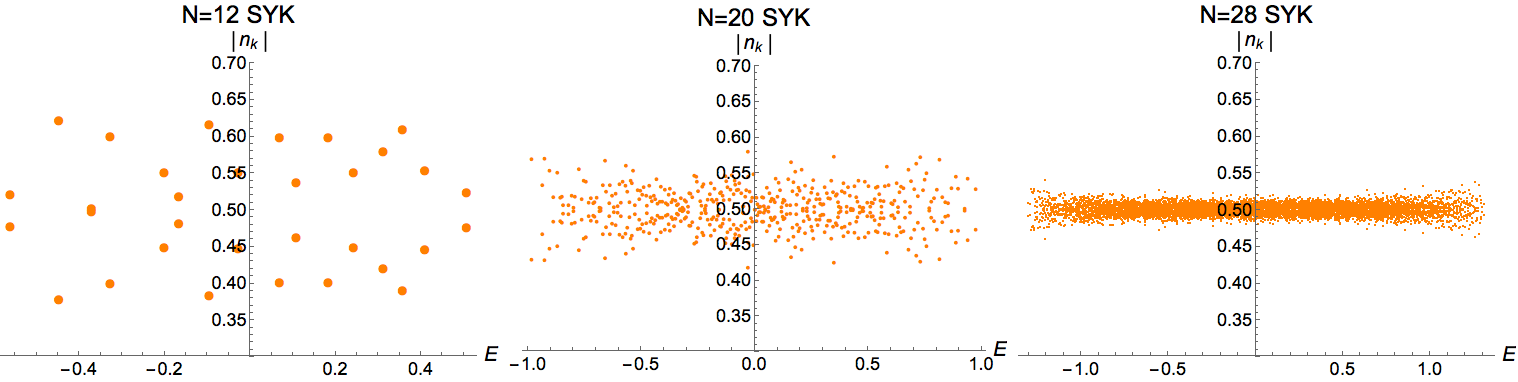}\\
\includegraphics[width=0.92\textwidth]{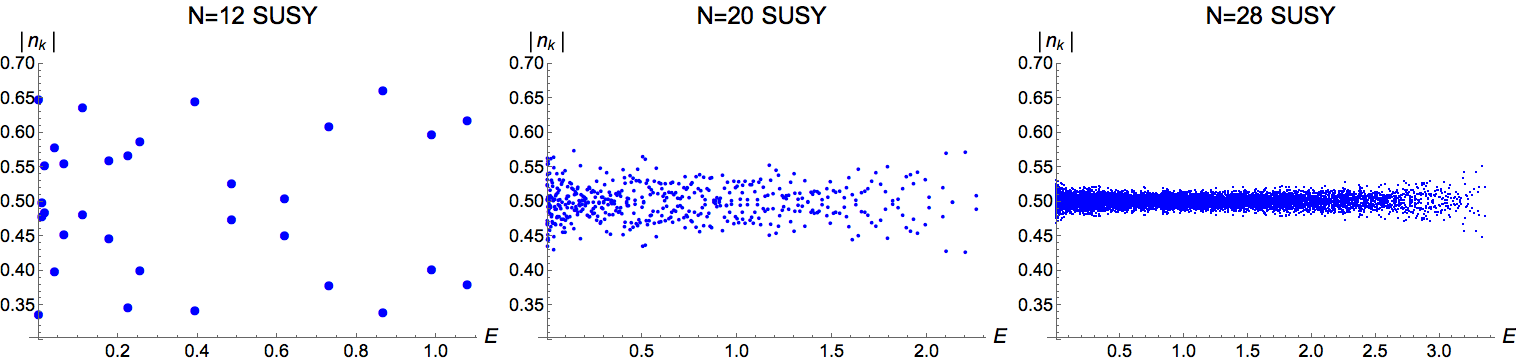}
\caption{Single realization of diagonal matrix elements for a single number operator $n_k$ in the SYK and supersymmetric SYK model (energy eigenvalues vs $|n_k|$.). We consider $N=$ 12, 20, 28 and observe the diagonal terms fluctuating around $1/2$.} \label{1realizenontri}
\end{figure}

\begin{figure}
\centering
\includegraphics[width=0.8\textwidth]{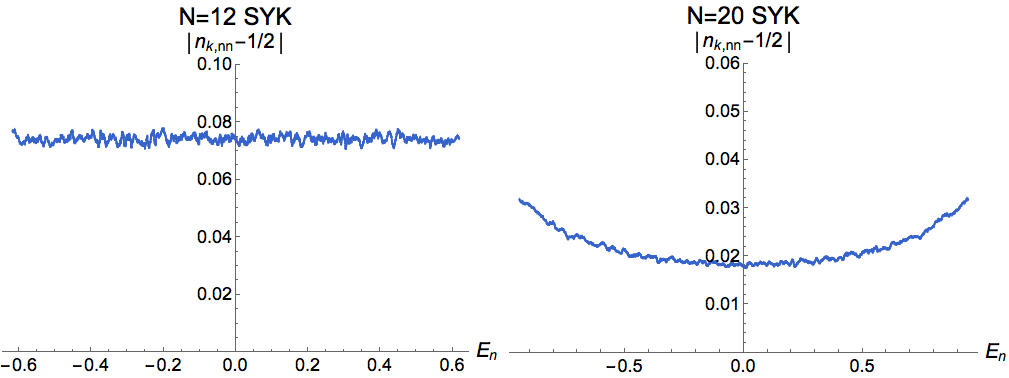}
\includegraphics[width=0.8\textwidth]{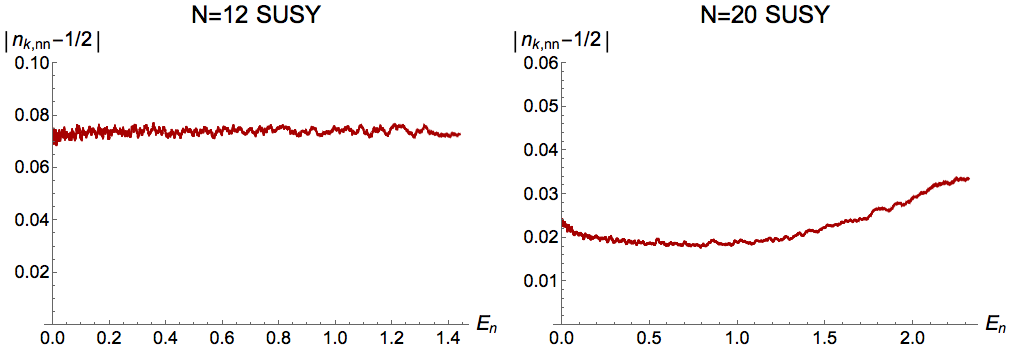}
\caption{Multiple realization of diagonal elements the number operator in SYK (above), supersymmetric SYK (below), taking $20000$ realizations for $N=12$ and $1000$ realizations for $N=20$. The plots are given by the energy eigenvalue dependence of the absolute difference between the diagonal elements and $1/2$, taking a nearest-neighbour moving average to smooth the curve.} \label{multirealizediag}
\end{figure}

One can see that for larger $N$, and with substantially more eigenvalues, the average deviation from $1/2$ should be a function of the eigenvalues and need not be a constant function. Thus we plot the smoothed deviations from the microcanonical value $1/2$, for $N=12$ and $N=20$ with multiple realizations of disorder, in Figure \ref{multirealizediag}. One can see that at lower $N$ (the $N=12$ case), the average deviation between diagonal terms and $1/2$ is roughly a constant, while for larger $N$ ($N=20$ here) there exists stronger fluctuations at larger energies, and more so at even larger $N$, for instance even in single realization in Figure \ref{1realizenontri}. This fact confirms our observation in Figure \ref{ratio}.

\subsection{Off-diagonal terms}
\begin{figure}
\centering
\subfigure{
\begin{minipage}{1.0\textwidth}
\includegraphics[width=1\textwidth]{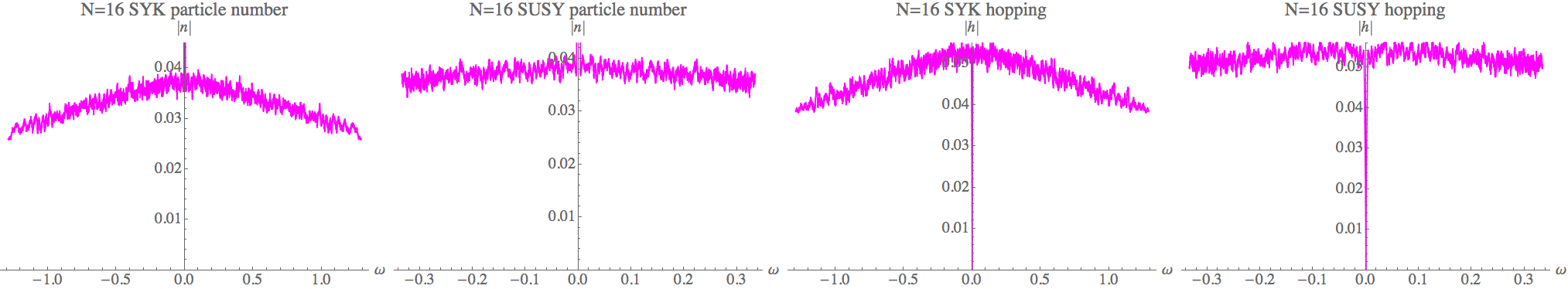}
\end{minipage}
}

\subfigure{
\begin{minipage}{1.0\textwidth}
\includegraphics[width=1\textwidth]{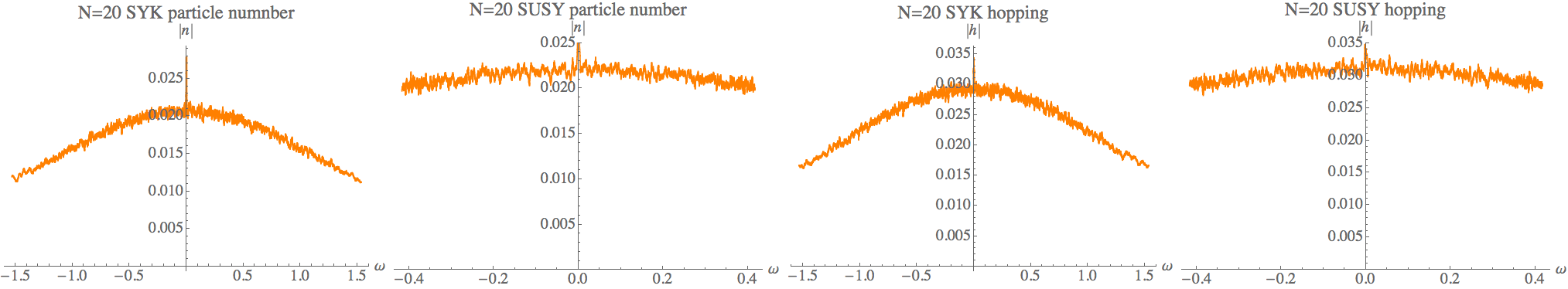}
\end{minipage}
}

\subfigure{
\begin{minipage}{1.0\textwidth}
\includegraphics[width=1\textwidth]{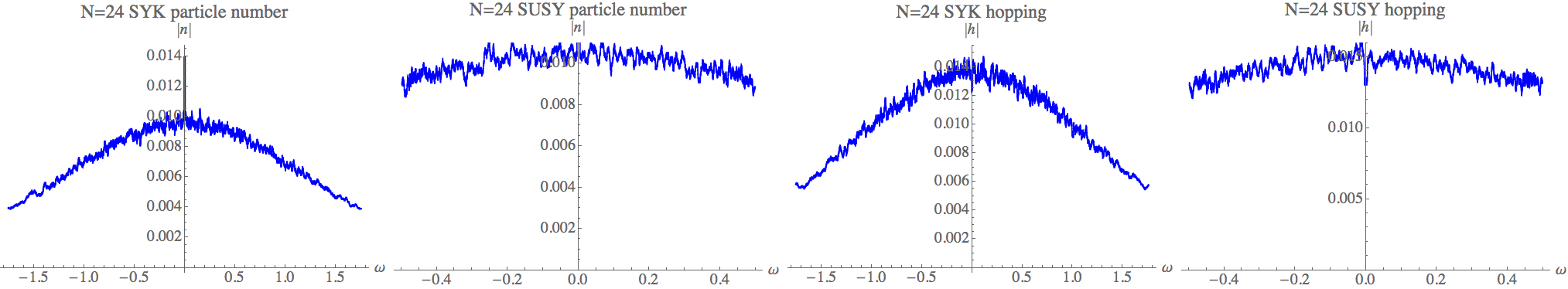}
\end{minipage}
}
\caption{Off-diagonal matrix element dependence on the energy difference $\omega = E_m-E_n$, at a fixed average energy $\bar E = (E_m+E_n)/2$, of the particle number operator and the hopping operator in SYK and supersymmetric SYK models, plotted for 1000 realizations of $N=16$ (top), $N=20$ (middle), and $N=24$ (bottom) in the two models. We consider fixed average energy $\bar{E}/N=0.02\pm0.001$ and take a moving average in $\omega$ (800 nearest neighbours) of the matrix elements. The peak at the center of the plots is due to the eigenstates becoming diagonal as $\omega \ra 0$.} \label{off}
\end{figure}

Another crucial numerical check of ETH is to consider the off-diagonal elements, looking at the statistics of the random fluctuations. We consider the single particle number and hopping operators in the SYK and supersymmetric models and in Figure \ref{off}, plot the statistical average of the off-diagonal elements as a function of the energy difference ($\omega = E_m-E_n$). One can observe some generic features in the numerics: First, the off-diagonal elements have larger variations in the middle of the curve, where the energy difference is small. Second, the difference between the bulk and the edge in the $\omega-|\mathcal{O}|$ distribution becomes sharper when $N$ becomes larger. Finally, in the supersymmetric model, the slope of the curve is smaller than the original SYK model. Some of these features are also seen in the complex fermion model \cite{Sonner17}. We also note that the decreasing variance of the fluctuations at larger energy differences (large $\omega$) is a universal feature in chaotic systems satisfying ETH, and the averaged decay of fluctuations at larger $\omega$ sets the scale of the Thouless energy and the diffusion time.

\subsection{Gaussianity of fluctuations}
The fluctuation term $R_{mn}$ in the off-diagonal contribution is often assumed to be Gaussian (see for instance, \cite{SrednickiETH}), which has been checked numerically in disordered models \cite{ETHgauss}. Here we will verify that the Gaussian condition is satisfied for both SYK and supersymmetric SYK models in each parity sector. To start, one can directly compute the distribution of off-diagonal terms from multiple realizations of disorder parameters for fixed energy average $\bar{E}$ and difference $\omega$ appearing in the spectrum. Generically, the distribution of $R_{mn}$ should be complex, and an overall complex phases do not change the distribution of $R_{mn}$ (see Appendix \ref{app:UI} a discussion of this). However, in the case of even $N_d$ we are considering, the only exception is $N \text{ mod }8=0$ case, which corresponds to real representation in random matrix theory classification (see \cite{You16,BHRMT16,Li:2017hdt} for reference), for both SYK and supersymmetric SYK models. In this case, one can use the real representation of Clifford algebra to ensure reality of the eigenvectors.

\begin{figure}
\centering
\includegraphics[width=0.92\textwidth]{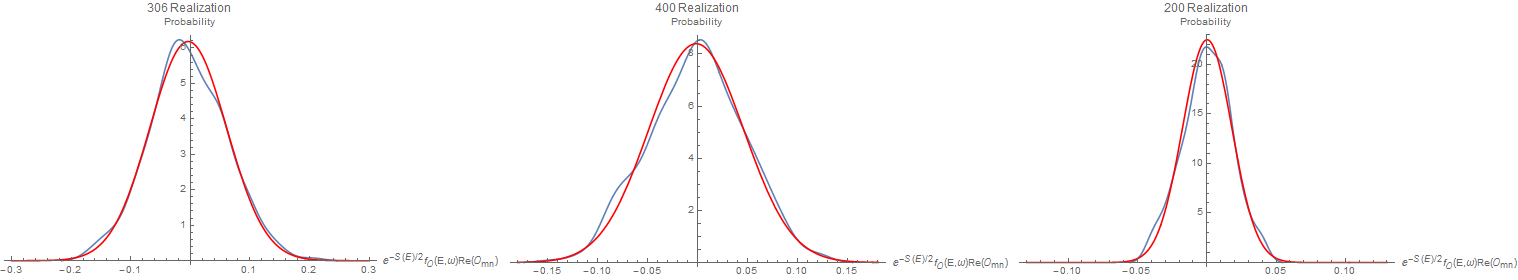}\\
\includegraphics[width=0.92\textwidth]{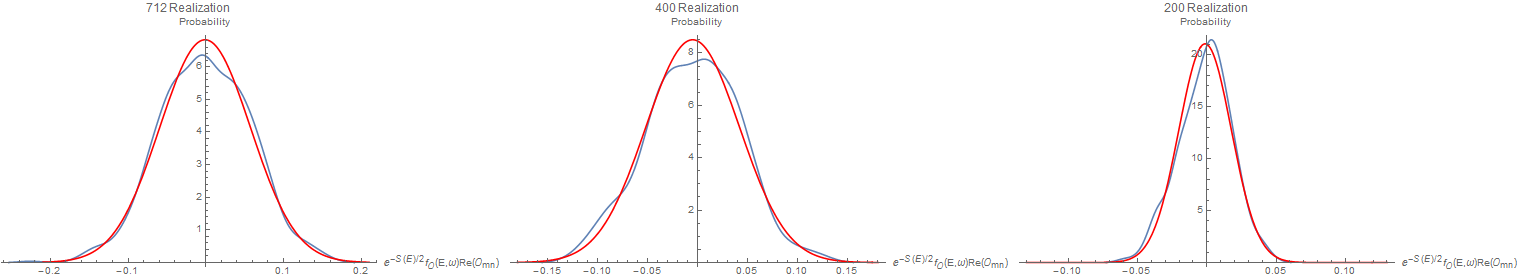}
\caption{Gaussianity of fluctations for fixed energy eigenvalues. For $N=$ 12, 16, 20 (left, middle, right) we check multiple disorder realizations for SYK (above) and supersymmetric SYK (below) models for fixed energies $\bar{E}=0.15$ and $\omega=0.1$ (i.e. $E_m=0.2$ and $E_n=0.1$).  The exact Gaussian distribution with the same expectation value and variance is plotted (red curves) alongside the smoothed statistical results (blue curves). Deviation from the Gaussian distribution is likely due in part to the finite sample size and finite $N$, and also a numerical artifact of smoothing.} \label{gauss}
\end{figure}

In Figure \ref{gauss} we check the Gaussian distribution of $R_{mn}$ for fixed energies. Those numerical results clearly show the Gaussianity of these fluctuations, while the numerical non-Gaussianity may due to the finiteness of sample numbers, the finiteness of $N$, and also the numerical artifacts when smoothing the distribution curves.

\subsection{Pure state thermalization}
There is another related statement of thermalization, which has a clear geometric interpretation in the SYK model, regarding the thermalization of the correlators under the Euclidean evolution of a certain set of pure state \cite{Kourkoulou:2017zaj}. The statement is given as the following: consider an arbitrary eigenstate of $S_k$, namely one of the $\ket{B_s}$ states we discussed previously, and define a low energy projection with the following Euclidean evolution
\begin{equation}
\ket{B_s(\ell )} =e^{-\ell H} \ket{B_s}\,.
\end{equation}
Given that the states $\ket{B_s}$ form a basis, we can take thermal averages as
\begin{equation}
\sum_s \braket{B_s (\ell) | \op | B_s (\ell)} = \Tr \big( e^{-\beta H} \op\big)\,,
\end{equation}
with $\beta=2\ell$. Moreover, a single pure state $B$ is also very useful to simulate the thermal state in the SYK model, namely the correlators are thermalized. Inserting the identity into the above expression, each term in the sum is equal in the large $N$ limit \cite{Kourkoulou:2017zaj}, meaning
\begin{align}\label{eq:rela}
\left\langle  {{B}} \right|{{e}^{-\beta H}}\left| {{B}} \right\rangle ={{2}^{-N/2}}Z(\beta )\,.
\end{align}
Moreover, we can define
\begin{align}\label{eq:cor}
  & {{G}^{\text{diag}}}({{\tau }_{1}},{{\tau }_{2}})=\frac{1}{\left\langle {{B}}(\ell )|{{B}}(\ell ) \right\rangle }\langle {{B}}(\ell )|{{\psi }^{i}}({{\tau }_{1}}){{\psi }^{i}}({{\tau }_{2}})\left| {{B}}(\ell ) \right\rangle \approx {{G}_{\beta }}({{\tau }_{1}}-{{\tau }_{2}})\nonumber\\
 & {{G}^{\text{off}}}({{\tau }_{1}},{{\tau }_{2}})=\frac{{{s}_{k}}}{\left\langle {{B}}(\ell )|{{B}}(\ell ) \right\rangle }\langle {{B}}(\ell )|{{\psi }^{2k-1}}({{\tau }_{1}}){{\psi }^{2k}}({{\tau }_{2}})\left| {{B}}(\ell ) \right\rangle \approx -2i{{G}_{\beta }}({{\tau }_{1}}){{G}_{\beta }}({{\tau }_{2}})\,,
\end{align}
where
\begin{align}
{{G}_{\beta }}(\tau )=\frac{1}{N}\sum\limits_{i}{\frac{\text{Tr}\left( {{e}^{-\beta H}}{{\psi }^{i}}(\tau ){{\psi }^{i}}(0) \right)}{\text{Tr}\left( {{e}^{-\beta H}} \right)}}\,.
\end{align}
These expressions are valid in the large $N$ limit $\beta J\ll N$. In the limit $\tau\to 0$, each term in the sum of the thermal correlator gives $\vev{\psi^i(0)\psi^i(0)}_\beta/Z = 1/2$. The thermalization of these pure state is a consequence of symmetry of the replica symmetric disordered model (a subgroup of $O(N)$ termed the flip group). Considering that $\mathcal{N}=1$ supersymmetry only changes the distribution of the Hamiltonian (up to a constant energy shift), the flip group argument is still valid. As a result, we expect that the correlators in the low energy pure are still thermal.

\begin{figure}
\centering
\includegraphics[width=0.5\textwidth]{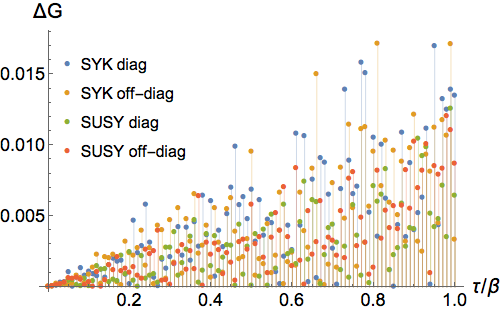}
\caption{Pure state thermalization: we consider $N=12$ models for both SYK and the supersymmetric model and compute the relative difference between the correlation function in the pure state (LHS of Eq.~\ref{eq:cor}) and in the thermal state (RHS of Eq.~\ref{eq:cor}). We fix $\beta=2$ and consider 500 realizations. The small relative difference supports the pure state thermalization in both models.} \label{pure_therm}
\end{figure}

We confirmed these expectations with numerics, comparing the pure state inner product and the partition function (namely, verifying Eq.~\eqref{eq:rela}) and comparing the diagonal and off-diagonal correlators with thermal ones (namely, verifying Eq.~\eqref{eq:cor}), and find very good numerical agreement. As an example, the comparison with the thermal correlators is shown in Figure \ref{pure_therm}. This indicates thermalization of low energy projected pure states.

\section{Discussion and Conclusion}
\label{sec:con}
In this paper we numerically investigated the thermalization of energy eigenstates in the Majorana SYK and supersymmetric SYK models. We focused primarily on conventional few-body operators, the single particle number and the hopping operators, to verify that these models satisfy ETH. Our results include the direct verification of ETH by checking the diagonal dominance and the relative difference between microcanonical ensembles, the statistics of the diagonal terms and the off-diagonal terms, and verifying the distribution of off-diagonal fluctuations. The numerical evidence indicates that SYK and its $\mathcal{N}=1$ supersymmetric extensions satisfy ETH as defined in Eq.~\eqref{eq:ETH}.

Although ETH applies in a broad class of quantum systems, the statement that ETH holds in the SYK model confirms a basic expectation of a system with a gravitational dual: correlation functions in energy eigenstates of look thermal, as anticipated from local bulk effective field theory and in line with results from 2d CFT \cite{ThermalBlocks}. There is a direct relation \cite{Sonner17} between the thermalization of eigenstates and thermal correlators in low-energy pure states constructed in \cite{Kourkoulou:2017zaj}. In SYK we checked this thermalization for the pure state correlators in the large $N$ limit. As these pure states are suggestive of two-dimensional black hole microstates, operators related to these pure states could be used to study transversable wormholes \cite{Gao:2016bin,Maldacena:2017axo}.

Beyond making statements about black hole thermalization in a putative gravitational dual, the claim that SYK satisfies ETH is interesting from a quantum information perspective. The frame potential for an ensemble of unitaries $\CE$ is a measure of how close that ensemble is to forming a unitary $k$-design \cite{Scott08}, \ie reproducing the first $k$ moments of the Haar ensemble. Building on this, \cite{ChaosRMT} defined the notion of $k$-invariance as a measure of how scrambled a system is under chaotic time-evolution, giving rise to a random matrix description. If a system achieves $k$-invariance, the eigenstates may be treated as random vectors. For a system satisfying ETH, the late-time dynamics should be $k$-invariant and, in this precise sense, random. Given recent information-theoretic approaches to studying chaos and complexity \cite{ChaosChannels,ChaosDesign,ChaosRMT}, it would be nice to understand the precise role ETH plays in this framework and whether it is a necessary condition for information scrambling or chaotic decay of correlation function. Moreover, evidence that SYK satisfies ETH hints at some error-correcting properties. As the condition for correctability of an approximate quantum error correcting code is satisfied by finite energy density eigenstates of systems which obey ETH \cite{ETH_QEC}, an analytic understanding of ETH in the SYK model might elucidate its underlying code properties.

Generically, one expects ETH to hold for few-body physical operators; but the statement might be more subtle for non-local systems. It would also be interesting to investigate eigenstate thermalization in generalizations of SYK with spatial locality \cite{Gu:2016oyy} and in tensor models \cite{Gurau:2010ba,Witten:2016iux,Klebanov:2016xxf,Krishnan:2016bvg}, to understand the role all-to-all interactions and quenched disorder play in the thermal structure of eigenstates. There are more fundamental subtleties about which energy windows and for which operators ETH actually applies. Recent progress has been made in addressing these issues, both in the context of chaotic many-body systems \cite{Grover15,Lu:2017tbo} as well as in 2d CFTs \cite{Lashkari:2016vgj,Basu:2017kzo}, alongside work towards generalizing ETH \cite{Dymarsky:2016aqv,Dymarsky:2017zoc}. Lastly, as ETH has not been proven to hold in any generic class of interacting models and analytic results of systems that satisfy ETH are few and far between, one might hope that SYK, a rare example of a strongly-coupled but solvable quantum mechanical model, affords the opportunity to prove ETH analytically. We leave these topics to future research.

\section*{Acknowledgments}
We thank Fernando Brand\~ao, Tarun Grover, Jenia Mozgunov, Burak {\c S}ahino{\u g}lu, David Simmons-Duffin, and Beni Yoshida, as well as the attendees of the Institute for Quantum Information (IQI) group meetings at Caltech, for valuable discussion and comments. We especially thank Yuan Xin for help on down-sampling the plots. NHJ acknowledges support from the Simons Foundation through the ``It from Qubit'' collaboration as well as from the Institute for Quantum Information and Matter (IQIM), an NSF Physics Frontiers Center (NSF Grant PHY-1125565) with support from the Gordon and Betty Moore Foundation (GBMF-2644). JL is supported by the U.S. Department of Energy, Office of Science, Office of High Energy Physics, under Award Number DE-SC0011632. YZ is supported by the graduate student program at the Perimeter Institute.

\appendix
\section{Constraints from Particle-Hole Symmetry}
\label{app:PHsym}
Both the SYK model and its supersymmetric extension admit a particle-hole symmetry \cite{You16,BHRMT16}
\begin{align}
P=K\prod\limits_{\alpha=1}^{N_d}{({{c}_{\alpha}}+c^\dagger_\alpha)}\,, \where [H,P]=0\,,
\end{align}
and where $K$ is the antiunitary complex conjugation operator. The particle-hole operator squares to $P^2 = (-1)^{\lfloor N/4 \rfloor}$ and acts on the fermions as
\begin{equation}
P c_k P = \theta c_k^\dagger\,, \quad P c_k^\dagger P = \theta c_k\,,\quad P\psi^i P = \theta \psi^i\,, \where \theta = (-1)^{N/2-1} P^2\,.
\end{equation}
We can consider the action of $P$ on an energy eigenstate $\ket n$. If $N/2$ is odd the the state is mapped from one charge parity sector to the other, exchanging even and odd. If $N/2$ is even $P$ maps each sector to itself, but if $N ~{\rm mod}~ 8 = 4$ then there is a degeneracy in each sector as $P^2 = -1$ implies $P\ket n$ must take us to a different eigenstate. If $N ~{\rm mod}~ 8 = 0$, there is no degeneracy and $P$ maps energy eigenstates to themselves (up to a phase).


This implies that when $N ~{\rm mod}~ 8 = 0$, we have an exact half-filling for energy eigenstates, namely that for any eigenstate $\ket n$ and any $k$, the number operator $n_k = c_k^\dagger c_k$ has
\begin{equation}
\braket{n | c_k^\dagger c_k|n} = \frac{1}{2}\,, \quad \text{or equivalently}\quad  \langle n|\psi ^i \psi ^j|n\rangle=0\,,
\label{eq:nvev}
\end{equation}
for any $i\neq j$. Recall that $n_k = (S_k+1)/2$ where $S_k = 2i \psi^{2k-1}\psi^{2k}$.

To show Eq.~\eqref{eq:nvev}, consider $\braket{n|n_k|n}$ and insert $P^2$
\begin{equation}
\braket{n|c_k^\dagger c_k|n} = (-1)^{\lfloor N/4\rfloor} \braket{ n|P^2 c_k^\dagger c_k| n} = (-1)^{\lfloor N/4\rfloor} \braket{ n|P (1-c_k^\dagger c_k) P| n}
\end{equation}
where we commute $P$ through as $P c_k^\dagger c_k = c_k c_k^\dagger P $. When $N~{\rm mod}~8 = 0$, $P$ then takes $\ket n$ to itself and we find the desired result.

Note that the above argument extends to any operator $\op_{2k}$ of the following form
\begin{equation}
\op_{2k}=\psi ^{i_1}\psi ^{i_2}\cdots \psi ^{i_{2k}}\,, \quad{\rm then}\quad \langle n| P \op_{2k} P|n\rangle=(-1)^k\langle n|\op_{2k}|n\rangle\,,
\end{equation}
for operators with no coincident indices $i_a \neq i_b$. Since we can assume that $i_a = a$, $\op_{2k}$ is proportional to a product of $2c_{\alpha }^{\dagger }{{c}_{\alpha }}-1$ with $\alpha$ running from 1 to $k$. In other words, for odd $k$,  $\op_{2k}$ always has zero trace on any energy eigenstates (degenerate or not), in particular, the diagonal terms of $\op_{2k}$ vanish when $N$ mod 8 $= 0$.

\begin{figure}
\centering
\includegraphics[width=0.92\textwidth]{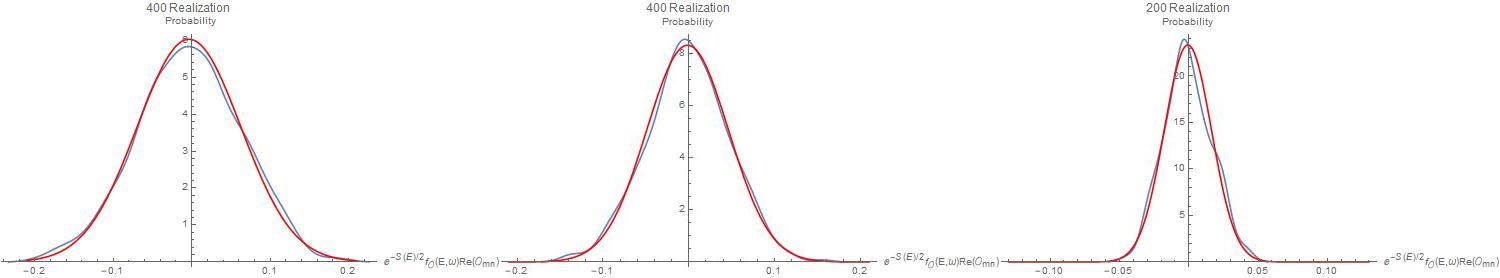}\\
\includegraphics[width=0.92\textwidth]{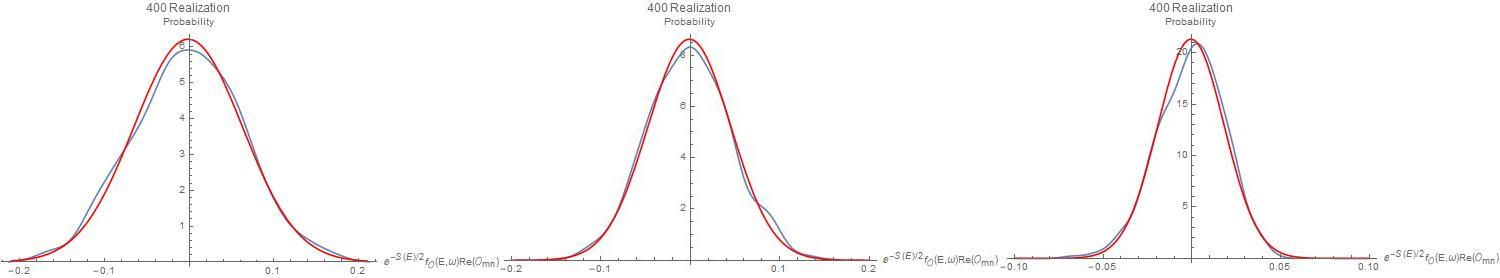}
\caption{Gaussian distribution for fixed energy eigenvalues. Again, for $N=$ 12, 16, 20 (left, middle, right) we check multiple disorder realizations for SYK and supersymmetric SYK models (up, down) for fixed energies $\bar{E}=0.15$ and $\omega=0.1$, or equivalently $E_m=0.2$ and $E_n=0.1$. The realization scheme is the same as the previous one, but we average phases over a uniform distribution.} \label{gauss_modified}
\end{figure}

\section{Unitary Averaging of the Distribution}
\label{app:UI}
Considering the ETH ansatz for the SYK model (and its supersymmetric generalizations),
\begin{align}
\vev{m|\op|n} =\bar{\mathcal{O}}(\bar{E}){{\delta }_{mn}}+{{e}^{-S(\bar{E})/2}}{{f}_{\mathcal{O}}}(\bar{E},\omega ){{R}_{mn}}\,,
\end{align}
a natural question then arises: if we want to look at the statistics of the off-diagonal random variables $R_{mn}$, how should we make sense of eigenstates $|m\rangle$ and $|n\rangle$ when doing numerics by exact diagonalization? More specifically, as we are averaging over random Hamiltonians, each realization comes with a choice of eigenstate up to a nonphysical phase. Here we shall propose a ETH for the SYK model and its supersymmetric generalizations by stating that non-physical phases should be averaged by a uniform distribution to remove all phase fluctuations. To be precise,
\begin{align}
\vev{m,\theta _m |\op| n,\theta _n} _ {\theta}&=\bar{\mathcal{O}} (\bar{E}){{\delta }_{mn}}+{{e}^{-S(\bar{E})/2}}{{f}_{\mathcal{O}}}(\bar{E},\omega ){{R}_{mn}}\nonumber \\
\left| k, \theta _k \right\rangle & =e^{i\theta _k}\left| k \right\rangle \text{, }k\in \{1,2,\cdots ,2^{N_d}\}\,,
\end{align}
where $\theta _k$ are independently chosen from $0$ to $2\pi$ with uniform distribution, and subscript $\theta$ indicates an average over all $\theta _k$. We also assume that the choice of $|n\rangle$ are not random, i.e. for a set of coupling constants there is only one set of eigenstates with fixed phases. In the numerical realization using Mathematica, the assumption holds because eigenstates are generated with some definite algebraic procedure.

Here we remark that the the distribution of phase of $R_{mn}$ does not depend on the choice of the original (fixed) choice of the phase of $|n\rangle$, because different choice of phase correspond to a rotation in complex plane and averages out over the uniform distribution of $\theta _k$. In Figure \ref{gauss_modified} we check the Gaussian distribution of $R_{mn}$ for fixed energies. Compared to the unmodified distribution in Figure \ref{gauss}, the non-Gaussianity is suppressed.

\bibliographystyle{utphys}
\bibliography{ETH_SYK}

\end{document}